\documentclass[twocolumn]{apex}

\title{Local Measurement of Microwave Response with Local Tunneling Spectra Using Near Field Microwave Microscopy}

\author{Tadashi Machida$^{1, 2}$, Marat B. Gaifullin$^{3}$, Shuuichi Ooi$^{1}$, Takuya Kato$^{2}$, Hideaki Sakata$^{2}$, and Kazuto Hirata$^{1}$}
\inst{$^{1}$Superconducting Materials Center, National Institute for Materials Science, 1-2-1 Sengen, Tsukuba, Ibaraki 305-0047, Japan \\
$^{2}$Department of Physics, Tokyo University of Science, 1-3 Kagurazaka, Shinjuku-ku, Tokyo 162-8601, Japan \\
$^{3}$Department of Physics, Loughborough University, Leicestshire, LE11 3TU, U.K.
}
\abst{We have fabricated near-field scanning microwave microscope and measured the local microwave response and the local density-of-states (LDOS) in the area including the boundary between the Au deposited and the non-deposited region on highly-orientated pyrolytic graphite at a frequency of about 7.3 GHz. We have succeeded in measuring both the LDOS and the surface resistance. It can be observed that the surface resistance in Au deposited region with the metallic tunneling spectra is smaller than that in the non-deposited region with the U-shaped tunneling spectra. 
}

\begin{document}
\maketitle
Scanning tunneling microscope (STM) has been used in the scanning tunneling spectroscopy (STS) mode to investigate a variety of issues such as probing the local density-of-states (LDOS) within the superconducting vortex cores~\cite{Hess, Matsuba}, elucidating the inhomogeneous distribution of superconducting gap in high temperature superconductors~\cite{Machida}, studying the edge states in graphite sheet~\cite{Niimi}, capturing the LDOS in carbon nano-tubes~\cite{Hassanien}, and so on.
On the other hand, near-field scanning microwave microscope (NSMM) enables us to measure locally the microwave response (MWR) from a sample. For example, the local dielectric coefficient in (Ba, Sr)TiO$_\mathrm{3}$~\cite{Steinhauer_2} and the local surface resistance ($R_\mathrm{sur}$) in YBa$_\mathrm{2}$Cu$_\mathrm{3}$O$_\mathrm{7-\delta}$~\cite{Steinhauer_1} have been visualized by the NSMM at mm or $\mathrm{\mu}$m length scale. Recently, an STM-assisted NSMM has been designed~\cite{Imtiaz_1} to integrate the functionality of both the STM and the NSMM.

Imtiaz \textit{et al}. achieved the spatial resolution of less than 100~nm by using the STM-assisted NSMM~\cite{Imtiaz_1, Imtiaz_3}. They claimed that the spatial resolution of the MWR approaches the value of the radius of curvature at the end of the STM tip ($r_\mathrm{tip}$), when the distance between the tip and the sample is much less than $R_\mathrm{tip}$. Until now, the local $R_\mathrm{sur}$ measurements have been done for the materials with relatively high $R_\mathrm{sur}$ of more than the order of 10 $\Omega$/sq. by a sharp tip with $r_\mathrm{tip}$ of less than a few $\mathrm{\mu}$m~\cite{Imtiaz_1, Imtiaz_2, Imtiaz_3}.
When the $R_\mathrm{sur}$ is lower than a few $\Omega$/sq., the rate of change in the MWR with respect to the change in $R_\mathrm{sur}$ becomes very small using the sharp tip~\cite{Imtiaz_1, Imtiaz_2, Imtiaz_3}.
For this reason, it was difficult to measure the local variation in $R_\mathrm{sur}$ in low $R_\mathrm{sur}$ materials.
Even though an STM-assisted NSMM holds the capability to measure both the local $R_\mathrm{sur}$ and the LDOS,
the spatial variation of these quantities has not been compared yet. 

The measurement of LDOS and local $R_\mathrm{sur}$ can reveal how the local variation of DOS due to impurities, defects, grains, carrier segregations, and so on, affects the local resistance on nano-meter length scales.
It is quite important to understand the electronic states around the local disorders.
In this study, we have succeeded in obtaining the clear difference in the MWR and LDOS near the boundary.  
We show the measurements of LDOS and local $R_\mathrm{sur}$ in low $R_\mathrm{sur}$ materials.
As a suitable sample, we use the boundary between Au deposited and non-deposited region on highly-orientated pyrolytic graphite (HOPG) whose $R_\mathrm{sur}$ is the order of 0.1 $\Omega$/sq, because of the following three reasons: (i) the surfaces of HOPG and Au are stable so as to obtain good tunneling spectra, (ii) there is a clear difference between the surface structures, and, (iii) these have low surface resistances.
\begin{figure}
\begin{center}
\includegraphics[width=7.5cm]{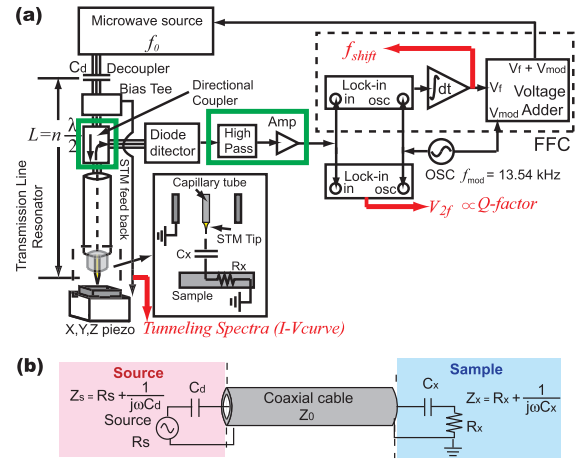}
\end{center}
\caption{(a) Schematic figure of the transmission line resonator based STM-assisted NSMM. The FFC keeps the microwave source locked onto one resonant frequency of the line resonator with the feed back time of about 100 $\mathrm{\mu}$s. The bias tee connects the STM feed back circuit to the center conductor of the coaxial line resonator. The four independent quantities can be measured as the function of the position: STM topography, resonator quality factor (Q), the frequency shift, and the tunneling spectra ($I-V$ curve or $dI/dV$ curve), as written by the red characters. The parts surrounded by the green boxes are differences from the design in ref. 8; (b) Schematic figure of the equivalent model in our resonator.}
\label{Fig1}
\end{figure}

The NSMM design used in the experiments is based on that in ref. 8. The transmission line resonator in the NSMM consists of the bias tee (IMMET 8810E), the directional coupler (HP87301D), and the coaxial cable whose impedance is 50 $\mathrm{\Omega}$, as shown in Fig. 1(a). One end of the resonator is connected to the home-made decoupling capacitor (labeled "Decoupler" in Fig. 1(a), $C_\mathrm{d}$ = 0.15 pF).  The other end terminates at an open-ended coaxial probe. This open-ended coaxial probe is just a piece of coaxial cable, in which a stainless steel capillary tube, holding the STM tip, replaces the center conductor. The bias tee permits us to use the same tip for both STM and microwave microscope. The reflected wave travels via the directional coupler to the diode detector (HP 8473C).
This detector produces a voltage signal, which is proportional to the power of the reflected wave from the resonator.
The output from this detector is sent to the two lock-in amplifiers referred at the external oscillator frequency $f_\mathrm{mod}$ (= 13.54kHz). The frequency-feedback-circuit (FFC) including one lock-in amplifier as marked out by dashed box in Fig. 1(a) keeps microwave source locked to a resonant frequency of the resonator. The voltage output from the FFC is proportional to the shifts in the resonant frequency ($\Delta$\textit{f}). In this case, this frequency rate is 10 MHz/V. This output voltage is added to the voltage oscillating at the frequency $f_\mathrm{mod}$ of the oscillator. The added voltage is used to modulate the source frequency at $f_\mathrm{mod}$. The deviation of the frequency modulation is corresponding to the amplitude of the oscillated voltage from the oscillator and is set to be 5 MHz in the measurement.  The detailed information with respect to the FFC can be seen in ref. 8.
The lock-in amplifier outside the dashed box in Fig. 1(a) picks up the signal at 2$f_\mathrm{mod}$ [labeled as $V_\mathrm{2f}$ in Fig. 1(a)], which gives a measure of the quality factor (Q) of the resonator.
In the apparatus, following two differences from the design in ref. 8 have been adopted to pick up a minute change of the signal [as shown by the green boxes in Fig. 1(a)]: (i) we put the directional coupler into the resonator to enhance the measured power of reflective wave from the sample, and (ii) we append the high-pass filter and the preamplifier to the system, to improve the signal noise ratio. The STM used was manufactured by UNISOKU. The STM feedback system can keep the tip-sample distance a few nm. This system allows us to measure simultaneously the four independent signals; the frequency shift ($\Delta$\textit{f}), the quality factor (\textit{Q}), the STM topography, and the local tunneling spectrum.

The resonator can be modeled as a resonant coaxial transmission line of the total length $L$ (=1.7114 m) with a capacitor $C_\mathrm{x}$ and a resistance $R_\mathrm{x}$ as shown in Fig. 1(b). In this model, resonant frequency \textit{f} and \textit{Q} are expressed by eqs. (1) and (2), respectively~\cite{Anlage_1}. 
\begin{eqnarray}
f_{(C_\mathrm{x}, R_\mathrm{x})}= \frac{1}{2\pi \sqrt{\epsilon_{r}\mu _{0} }}\Bigl(B_\mathrm{s}- \frac{\omega C_\mathrm{x}Z_\mathrm{0}}{1+\omega^2C_\mathrm{x}^2R_\mathrm{x}^2}\Bigr),
\end{eqnarray}
\begin{eqnarray}
Q_{(C_\mathrm{x}, R_\mathrm{x})} = \frac{B_\mathrm{s}(1+\omega^2C_\mathrm{x}^2R_\mathrm{x}^2)-\omega C_\mathrm{x}Z_\mathrm{0}}{A_\mathrm{s}(1+\omega^2C_\mathrm{x}^2R_\mathrm{x}^2)+\omega^2 C_\mathrm{x}^2Z_\mathrm{0}R_\mathrm{x}},
\end{eqnarray}
where $Z_\mathrm{0}$ is the impedance of coaxial cable (= 50 $\Omega$), and $\omega = 2\pi f$. 
Both quantities are dependent on the $C_\mathrm{x}$, corresponding to the tip-sample separation, and the sheet resistance of the sample $R_\mathrm{x}$.
$A_\mathrm{s}$ and $B_\mathrm{s}$ are constants related to the impedance mismatching at the position of the decoupling capacitor, as expressed below.
\begin{eqnarray}
A_\mathrm{s} = \alpha L +\frac{\omega^2C_\mathrm{d}^2Z_\mathrm{0}R_\mathrm{s}}{1+\omega^2C_\mathrm{d}^2R_\mathrm{s}^2},
\end{eqnarray}
\begin{eqnarray}
B_{s} = n\pi - \frac{\omega C_\mathrm{d}Z_\mathrm{0}}{1+\omega^2C_\mathrm{d}^2R_\mathrm{s}^2},
\end{eqnarray}
where $n$ is the mode number (= 118), $\alpha$ is the transmission line attenuation constant (0.49 Np/m), and $R_\mathrm{s}$ is the internal impedance of the source (= 50 $\mathrm{\Omega}$).
The local $C_\mathrm{x}$ and $R_\mathrm{x}$ values can be estimated from the locally measured resonant frequency and the quality factor at each location by using following two equations. 
\begin{eqnarray}
C_{x} = \frac{Q_\mathrm{n}^2(1-f_\mathrm{n})^2B_\mathrm{s}^2 + A_\mathrm{s}^2(f_\mathrm{n}-Q_\mathrm{n})^2}{Q_\mathrm{n}^2(1-f_\mathrm{n})^2B_\mathrm{s}Z_\mathrm{0}},
\end{eqnarray}
\begin{eqnarray}
R_{x} = \frac{Q_\mathrm{n}(f_\mathrm{n}-Q_\mathrm{n})A_\mathrm{s}Z_\mathrm{0}}{Q_\mathrm{n}^2(1-f_\mathrm{n})^2B_\mathrm{s}^2 + A_\mathrm{s}^2(f_\mathrm{n}-Q_\mathrm{n})^2},
\end{eqnarray}
where $f_\mathrm{n}$ and $Q_\mathrm{n}$ are the normalized resonant frequency and quality factor by the resonant frequency $f_\mathrm{0}$ (= 7.30445 GHz) and the quality factor $Q_\mathrm{0}$ without sample, respectively: $f_\mathrm{n} = f/f_\mathrm{0}$ and $Q_\mathrm{n} = Q/Q_\mathrm{0} \sim V_\mathrm{2f}/V_\mathrm{2f0}$ ($V_\mathrm{2f0}$ = 5.349 V).

\begin{figure}
\begin{center}
\includegraphics[width=6.5cm]{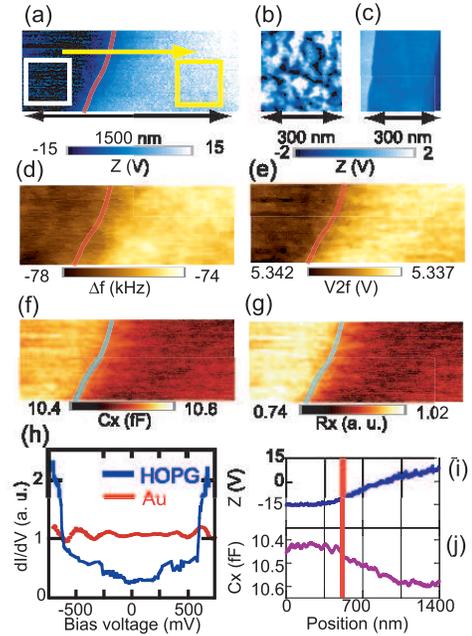}
\end{center}
\caption{(a) STM image of an area including the boundary between the Au deposited and non-deposited region on the surface of HOPG;  (b) and (c) fine STM images of the regions surrounded by the white and yellow squares in (a), respectively; (d) the frequency shift image and (e) the $V_\mathrm{2f}$ image in the same field of view of (a); (f) $C_\mathrm{x}$; (g) $C_\mathrm{x}$ and normalized $R_\mathrm{x}$ maps calculated from the data sets of (d) and (e) by using eqs. (5) and (6); (h) Tunneling spectra taken in Au deposited (red) and non-deposited region (blue); (i) and (j) the line profiles of the topography and local $C_\mathrm{x}$, respectively, along yellow arrow in (a). The red or light blue lines in each map and line profile are corresponding to the boundary between the Au deposited (right side) and non-deposited region (left side).}
\label{Fig2}
\end{figure}

To confirm the validity of the measurement of the LDOS and the local $R_\mathrm{sur}$, we used the boundary between the Au deposited and non-deposited region in the partially Au deposited HOPG, where the thickness of the deposited Au was 40 nm. The measurements were performed in air at room temperature. 
The probe tip used is mechanically etched Pt-Ir tip whose $r_\mathrm{tip}$ is about a few hundred nm. Figure 2(a) shows an STM image in the area including the boundary (red line) between the Au deposited (right side) and non-deposited region on HOPG (left side). 
On the left side from the boundary, the topography is almost flat. As the tip moves towards the right side from the boundary, the altitude gradually increases, as shown in Fig. 2(i). Figures 2(b) and 2(c) are the fine images of the regions surrounded by the yellow and white squares in Fig. 2(a), respectively. In Fig. 2(b), there are granular structures as expected on the Au surface~\cite{Imtiaz_1}. On the other hand, in Fig. 2(c), the step structures and the flat terraces as expected on the HOPG~\cite{Niimi} can be observed. Figures 2(d) and 2(e) show the $\Delta f$ and $V_\mathrm{2f}$ maps in the same field of view of Fig. 2(a). In spite of the low $R_\mathrm{sur}$ in the sample, these maps display the clear contrasts which are similar to the topography. As shown in Figs. 2(f) and 2(g), the local $C_\mathrm{x}$ and $R_\mathrm{x}$ were calculated from $\Delta f$ and $V_\mathrm{2f}$ maps by using eqs. (5) and (6).
In these maps, one can see the inverse contrast against the topography. The $C_\mathrm{x}$ start to decrease as going to the right side from the boundary, as shown in Fig. 2(j). When the STM is in the constant-current-mode, the probe follows the topography of the sample as the distance between the tip and the sample is kept to be constant. 
As the tip goes from the left to the right side, the capacitance becomes small as shown in Fig. 3(a), which is the first order approximation based on the models in refs. 8 and 9.  
As a result, it can be expected that the spatial variation of $C_\mathrm{x}$ shows the reverse change from the surface topography, during the scanning operation. 
Thus, the contrast in the $C_\mathrm{x}$ map reflects the surface topography.
\begin{figure}
\begin{center}
\includegraphics[width=6.5cm]{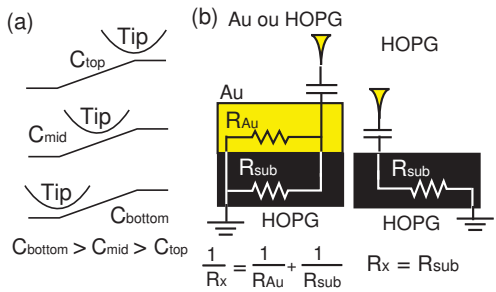}
\end{center}
\caption{(a) A first order approximation of the capacitance between tip and sample, when the tip is above the top (upper), mid-point (middle), and bottom (lower) on the slope; (b) Schematic figures of the resistances included in our resonator in our, based on the lamped elemented model in ref. 11. The left and right figures show the models when the tip is mounted on the Au deposited and non-deposited region, respectively.
}
\label{Fig3}
\end{figure}
The values in the $R_\mathrm{x}$ map have been scaled by the averaged value in the HOPG region, because it is difficult to estimate the absolute value of $R_\mathrm{x}$ due to the strong dependence of the $R_\mathrm{x}$ on the tip geometry~\cite{Imtiaz_1, Imtiaz_3}. 
The $R_\mathrm{x}$ in the Au region is smaller than that in HOPG region, as shown in Fig. 2(g).
The measured surface resistances $R_\mathrm{x}$ in Au deposited region should be considered as the combined resistance of the $R_\mathrm{Au}$ (= $\rho _\mathrm{Au}/t$, $t$ is the thickness of Au) and $R_\mathrm{sub}$ (=$\rho _\mathrm{HOPG}/\delta$, $\delta$ is the skin depth of HOPG whose value is about 4 $\mu$m) which is added in parallel with $R_\mathrm{Au}$, as shown in Fig. 3(b): 1/$R_\mathrm{x}$ =1/$R_\mathrm{Au}$ + 1/$R_\mathrm{sub}$ .
On the other hand, in non-deposited region, the $R_\mathrm{x}$ should contain only $R_\mathrm{sub}$: $R_\mathrm{x}$ = $R_\mathrm{sub}$.
Based on this perspective, the $R_\mathrm{x}$ in Au deposited region is about 20\% small against the $R_\mathrm{x}$ in non-deposited region, which is the nearly same as that in the observation.

In addition to the spatial variation of $R_\mathrm{x}$, the clear differences in tunneling spectra between Au deposited region and non-deposited region, which are proportional to the LDOS, can be observed, as shown in Fig. 2(h).
The LDOS in Au deposited region hardly depends on the energy as expected for the LDOS in Au.
On the other hand, U-shaped tunneling spectrum has been obtained in non-deposited region.
This U-shaped spectrum has been also observed in the STS experiments using normal STM~\cite{Niimi}.

The results in this study indicate the capability of the system to measure the local $R_\mathrm{sur}$ and the LDOS at nano-scale.
Consequently, the STM-assisted NSMM can provide the relation between LDOS and the local $R_\mathrm{sur}$ around an object which disturbs the local electronic properties.
Additionally, it is conceivable that the apparatus is suitable for the local measurement of the electron spin resonance in magnetic materials, the penetration depth measurements in superconductors, and the resistance in nano-particles and nano-wires.

In summary, we have demonstrated the local measurement of the tunneling spectra and the sheet resistance in the area including the boundary between the Au deposited and non-deposited region on HOPG by using the NSMM. 
Both the clear contrast in the local sheet resistance and the difference of tunneling spectra has been observed around the boundary simultaneously.
Our STM-assisted NSMM can measure the local density-of-states and local surface resistance in nano-scale.
This system is useful tool to study the effect of a local disorder on the electrodynamics and to measure the local properties in nano-materials.

\end{document}